\documentclass[aps, twocolumn,showpacs,showkeys]{revtex4}
\usepackage[]{graphicx}
\usepackage{amssymb}
\usepackage{bm}

\newcommand{\eq}[1]{Eq.~(\ref{#1})}
\newcommand{\fig}[1]{Fig.~\ref{#1}}         
\newcommand{\iotabar}{{\mbox{$\iota\!\!$-}}}

\newcommand{\mmax}{m_{\mathrm{max}}}

\begin{document}
\setcounter{page}{1}
\title[ ]{Quantum chaos? Genericity and nongenericity in the MHD spectrum of nonaxisymmetric toroidal plasmas}
\author{R.L. Dewar}
\email{robert.dewar@anu.edu.au}
\affiliation{%
 Department of Theoretical 
 Physics, Research School of Physical Sciences and Engineering,\\
 The Australian National University, ACT 0200, Australia}
\author{C. N\"uhrenberg}
\affiliation{%
Max-Planck-Institut f\"ur Plasma Physik,
Teilinstitut Greifswald, D-17491 Germany}
\author{T. Tatsuno}%
\affiliation{%
Center for Scientific Computation and Mathematical Modeling,
University of Maryland,
College Park, MD 20742-3289, USA
}%
\author{B.F. McMillan}
\affiliation{%
 Centre de Recherche en Physique des Plasmas,
 Ecole Polytechnique F\'ed\'erale de Lausanne,\\
 CH-1015 Lausanne,
 Switzerland}
\author{B.G. Kenny}
\affiliation{%
 Department of Theoretical 
 Physics, Research School of Physical Sciences and Engineering,\\
 The Australian National University, ACT 0200, Australia}

\date[]{Received ...}

\begin{abstract}
The eigenmode spectrum is a fundamental starting point for the
analysis of plasma stability and the onset of turbulence, but the
characterization of the spectrum even for the simplest plasma model,
ideal magnetohydrodynamics (MHD), is not fully understood. This is especially true in configurations with no continuous geometric symmetry, such as a real tokamak when the discrete nature of the external magnetic field coils is taken into account, or the alternative fusion concept, the stellarator, where axisymmetry is deliberately broken to provide a nonzero winding number (rotational transform) on each invariant torus of the magnetic field line dynamics (assumed for present purposes to be an integrable Hamiltonian system).

Quantum (wave) chaos theory provides tools for characterizing the spectrum
statistically, from the regular spectrum of the separable case
(integrable semiclassical dynamics) to that where the semiclassical
ray dynamics is so chaotic that no simple classification of the
individual eigenvalues is possible (quantum chaos). 

The MHD spectrum exhibits certain nongeneric properties, which we show, using a toy model, to be
understable from the number-theoretic properties of the asymptotic spectrum in the limit of large
toroidal and poloidal mode (quantum) numbers when only a single radial mode number is retained.

Much more realistically, using the ideal MHD code CAS3D, we have constructed a data set of several hundred growth-rate eigenvalues for an interchange-unstable three-dimensional
stellarator equilibrium with a rather flat, nonmonotonic rotational
transform profile. Statistical analysis of eigenvalue spacings shows evidence
of generic quantum chaos, which we attribute to the mixing effect of having a large number of
radial mode numbers.
\end{abstract}

\pacs{52.35.Bj,
          05.45.Mt 
               }
\keywords{Fusion plasma, Stellarator, Interchange instability, Suydam, Mercier, Essential Spectrum,
Quantum Chaos, Farey tree}

\maketitle

\section{Introduction}

The tokamak and stellarator fusion concepts both seek to contain a plasma in a toroidal
magnetic field, but to a good approximation the tokamak field is axisymmetric. Thus the
study of the spectrum of normal modes of small oscillations about equilibrium is simplified
by the existence of an ignorable coordinate. The stellarator class of device,
on the other hand, is inherently nonaxisymmetric and the lack of a continuous symmetry means 
there are no ``good quantum numbers'' to characterise the spectrum.

This makes the numerical computation of the spectrum a challenging task, but
numerical matrix eigenvalue programs, such as the three-dimensional
TERPSICHORE \cite{anderson_etal90} and CAS3D \cite{schwab93}
codes, are routinely used to assess the ideal magnetohydrodynamic
(MHD) stability of proposed
fusion-relevant experiments.  An example
is the design of the 5-fold-symmetric Wendelstein 7-X (W7--X)
stellarator, where CAS3D was used \cite{nuehrenberg96} to study a
number of different cases.

\begin{figure}[tbp]
\begin{center}
\begin{tabular}{cc}
      \includegraphics[scale=0.5]{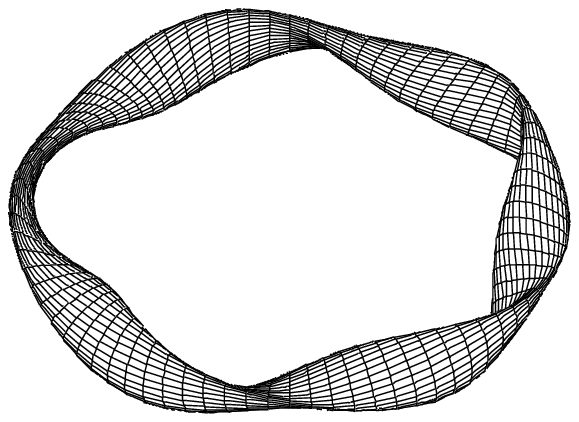}
    & \includegraphics[scale=0.3]{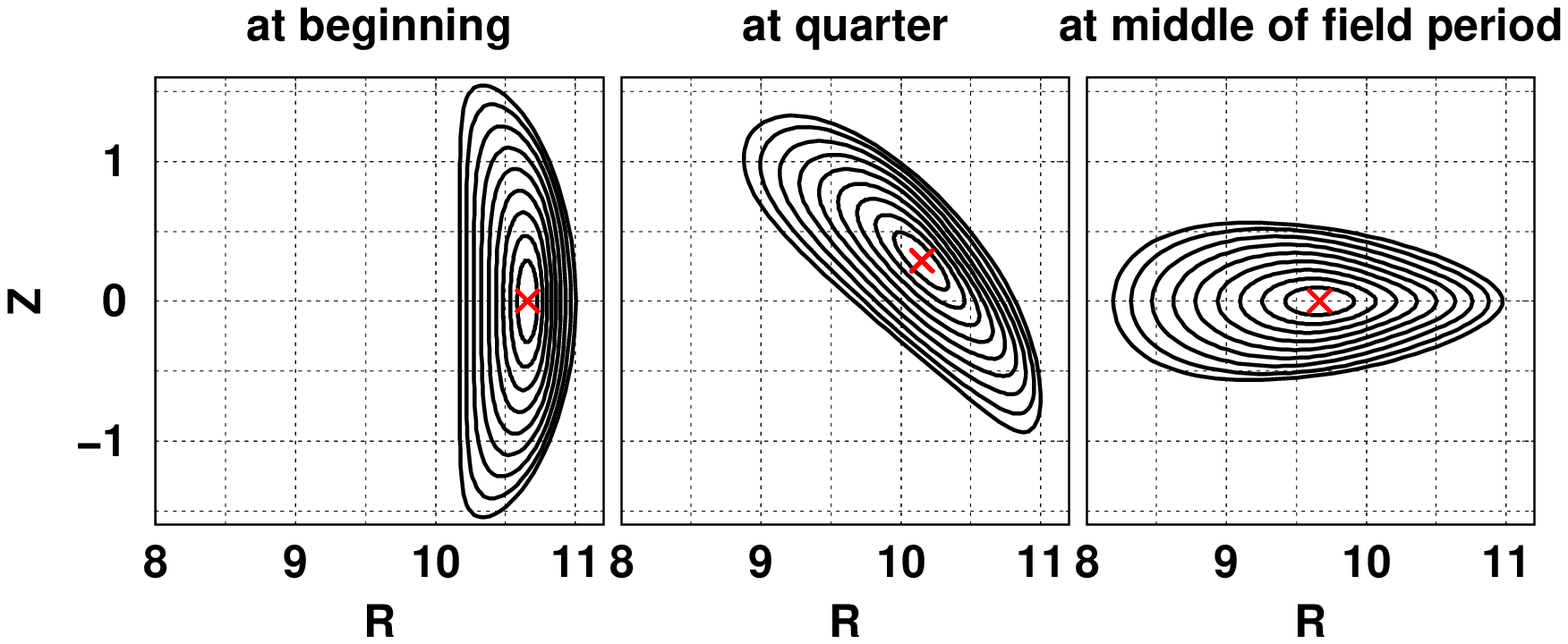}  \\
\end{tabular}
\end{center}
\caption{Left: plasma boundary of a 5-periodic toroidal equilibrium
geometrically related to W7--X configurations (at $\langle\beta\rangle
= 0.05$, VMEC calculation).  Right three frames: vertical
cross-sections of the configuration at three different toroidal angles
within a field period, with $Z$ the vertical coordinate and $R$ the
distance from the $Z$ axis. The plasma boundary has been scaled such that
the minor radius of the torus $a\approx 1$ and the major radius 
equals the aspect ratio $R/a=A$.}
    \label{fig:W7X}
\end{figure}

The configurations foreseen to be studied experimentally in W7-X are 
MHD stable,
but in the
present paper we are concerned with an unstable case from this study, 
a high-mirror-ratio, high-rotational transform equilibrium
(\fig{fig:W7X}). 
Due to its less pronounced shaping this case is quite unstable,
which contrasts with the properties of genuine W7-X configurations.
The three-dimensional nature of the equilibrium breaks
all continuous symmetries, coupling both poloidal ($m$) and toroidal
($n$) Fourier harmonics and thus precluding separation of variables
and simple classification of the eigenvalues.

These eigenvalues, $\omega^2 \equiv -\gamma^2$, are real due to the
self-adjointness \cite{bernstein_etal58} of the linearized force and
kinetic energy operators in ideal MHD linearized about a static
equilibrium.  This is analogous to the Hermitian nature of quantum
mechanics, so we might \emph{a priori} expect to be able to take over
mathematical techniques used in quantum mechanics.  Thus we study the
W7--X Mercier (interchange)-unstable case mentioned above using
statistical techniques from the theory of quantum chaos
\cite[eg]{haake01}.

This is of practical importance for numerical analysis of the
convergence of eigenvalue codes because, if the system is
quantum-chaotic, convergence of individual eigenvalues cannot be
expected and a statistical description must be used.  However, there
is a fundamental question as to whether the ideal MHD spectrum lies in
the same universality class as typical quantum mechanics cases.

This question has been addressed recently
\cite{dewar-nuehrenberg-tatsuno04,dewar_etal04} by
studying the interchange unstable spectrum in an effectively
cylindrical model of a stellarator.  In the cylindrical case the
eigenvalue problem is separable into three one-dimensional
eigenvalue problems, with radial, poloidal, and toroidal (axial)
quantum numbers $l$, $m$, and $n$, respectively.  If the spectrum
falls within the generic quantum chaos theory universality class for
integrable, non-chaotic systems \cite{berry-tabor77} then the probability
distribution function for the separation of neighboring eigenvalues is
a Poisson distribution.

However, this work indicates that the universality class depends on
the method of regularization (ie truncation of the countably infinite
set of ideal-MHD interchange growth rates): a smooth,
physically-motivated finite-Larmor-radius roll-off in the spectrum
appears to give the generic Poisson statistics for separable systems,
but a sharp truncation in $m$ and $n$ gives highly non-generic
statistics.  The latter case is less physical, but corresponds
closely to the practice in MHD eigenvalue studies of using a 
restricted $m,n$ basis set but a relatively fine mesh in the radial 
direction. 

A careful analysis of the spectrum of ideal-MHD
interchange modes in a separable cylindrical approximation
\cite{dewar_etal04} revealed non-generic behaviour of the spectral
statistics---a bimodal PDF, rather than the expected Poisson
distribution.  The non-genericity of this separable case indicates
that caution must be applied in applying conventional quantum chaos
theory in non-separable geometries.

The study \cite{dewar_etal04} indicated that the non-generic behaviour 
of ideal-MHD interchange modes was due to the peculiar feature of the 
dispersion relation for these modes that the eigenvalues in the 
short-wavelength limit depend only on the \emph{direction} of the wave 
vector, not on its magnitude. (This is unusual behaviour, but 
it is shared with internal gravity waves in geophysical fluid 
dynamics.) It was suggested in \cite{dewar_etal04} that the 
detailed features of the spectrum could be understood from the 
properties of Farey sequences.

In the present paper we discuss fictitious eigenvalues from a toy model, used to elucidate the
importance of number-theoretic effects, that illustrate how nongeneric the MHD eigenvalue spectrum can be if only one radial eigenmode is used.
Then we present the results of the quantum chaos
analysis of the W7--X case.



\begin{figure}[tbp]
    \begin{tabular}{cc}
		    \includegraphics[scale=0.5]{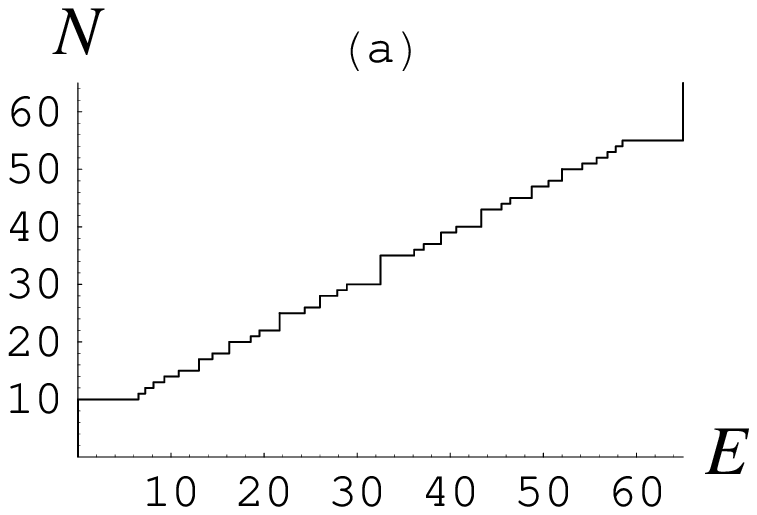} &
		    \includegraphics[scale=0.5]{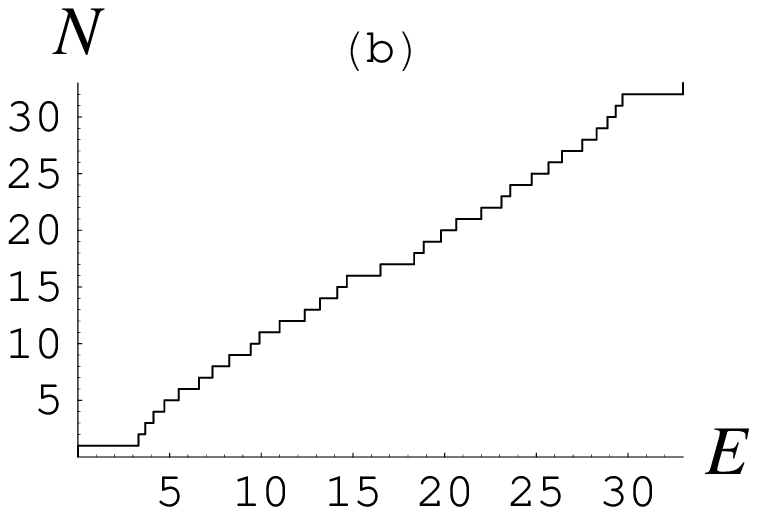} 
    \end{tabular}
    \caption{Number $N(E)$ of renormalized eigenvalues $E_{m,n}$ below a
    given value $E$ for $\mmax = 10$: (a) for the Hamiltonian $H =
    p_{\phi}/p_{\theta}$; (b) for the Farey sequence.}
    \label{fig:Devil}
\end{figure}

\section{Toy eigenvalue problem}
\label{sec:toy}

To gain insight into the nongeneric behavior found in the separable
case \cite{dewar_etal04} with only one radial eigenmode we study the energy
spectrum $\{{\cal E}_{n,m}\}$ for a ``toy'' quantum mechanical
Hamiltonian $H = p_{\phi}/p_{\theta}$ where the configuration space is
the 2-torus $\theta \in [0,2\pi)$, $\phi \in [0,2\pi)$ with periodic
boundary conditions.  In the semiclassical approximation we see that
$H$ depends only on the direction of {\bf p}, not its magnitude, as
for MHD interchange modes and internal gravity waves.

The eigenvalue problem is the time-independent Schr\"odinger equation,
$H\psi = {\cal E}\psi$, the eigenfunctions being $\exp [i(m\theta +
n\phi)]/4\pi^2$, where $m$ and $n$ are integers. The eigenvalues
${\cal E}_{n,m} = n/m$ ($m \neq 0)$.

Note the singular nature of the spectrum---it is discrete, yet
infinitely dense, the rationals being dense on the real line.  Also,
the spectrum is infinitely degenerate as eigenvalues are repeated
whenever $m$ and $n$ have a common factor.  Mathematically such a
spectrum, neither point nor continuous, belongs to the \emph{essential 
spectrum} \cite{hameiri85}.

In order to analyze this spectrum using standard quantum chaos
techniques we first regularize it by bounding the region of the $m,n$
lattice studied, and then allowing the bound to increase indefinitely.
Fortunately the PDF $P(s)$ is independent of the precise shape of the
bounding line when we follow standard practice \cite{haake01} in
renormalizing (unfolding) the energy levels to make the average
spacing unity.  Thus we adopt the simplest choice, taking the bounded
region to be the triangle $0 \leq n \leq m \leq \mmax$.  As the points
$(n,m)$ form a lattice in the plane with mean areal density of 1, we
can estimate the asymptotic, large-$\mmax$ behaviour of the number of
levels, $N_{\rm max}$, from the area of the bounding triangle: the $m$
axis, the line $n=m$ and the line $m=1$, which gives the ``Weyl
formula'' \cite{haake01} $N_{\rm max} \sim \mmax^2/2$.

The Farey sequence ${\cal F}(Q)$ is the set of all rational numbers
$p/q$ between 0 and 1, $1 < q \leq Q$, arranged in order of increasing
arithmetic size and keeping only mutually prime pairs of integers $p$ and
$q$.  Farey sequences are important in number theory
\cite{niven-zuckerman-montgomery91} and have application in various
dynamical systems problems, such as the theory of mode locking in
circle maps \cite{artuso-cvitanovic-kenny89}. They even have a
connection with the famous Riemann hypothesis \cite{augustin_etal01}.
 They are conveniently generated using the \emph{Mathematica} algorithm
 \cite{abbott92}
 \begin{verbatim}
 Farey[Q_] := Union[{0,1}, 
 \end{verbatim}
 
 \vspace{-10mm}
 \begin{verbatim}
 Flatten[Table[p/q,{q, Q},{p, q-1}]]].
\end{verbatim}
 
The list ${\cal G}(\mmax) \equiv \{{\cal E}_{n,m}\}$, sorted into a
non-decreasing sequence $\{{\cal E}_i|i = 1, 2,\ldots, N_{\rm max} \}$
is very similar to the Farey sequence ${\cal F}(\mmax)$ except for the
high degeneracy (multiplicity) of numerically identical levels,
especially when $n/m$ is a low-order rational.

Define the renormalized (unfolded) energy as $E_{n,m} \equiv N_{\rm
max}{\cal E}_{n,m}$.  The normalization by $N_{\mmax}$ ensures that
$E_{N_{\rm max}} = N_{\rm max}$, so the mean slope of the Devil's
staircase shown in \fig{fig:Devil}(a) is unity.  The large vertical
steps visible in \fig{fig:Devil}(a) are due to the high degeneracy at
low-order rational values of $n/m$.

\begin{figure}[tbp]
    \begin{tabular}{cc}
		    \includegraphics[scale=0.5]{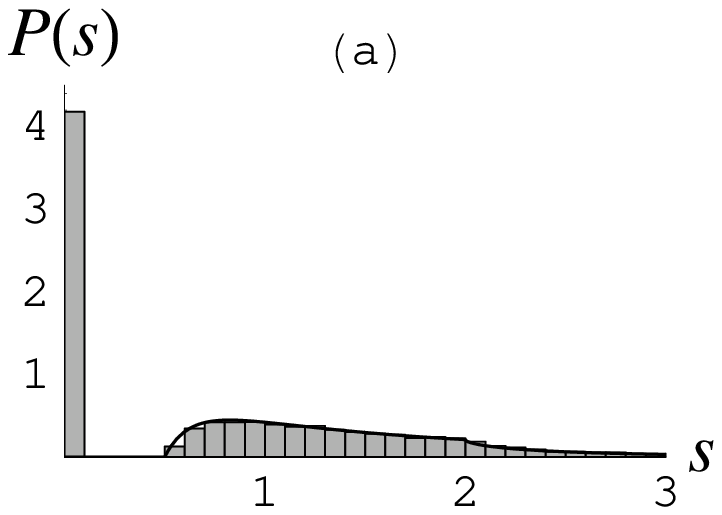} &
		    \includegraphics[scale=0.5]{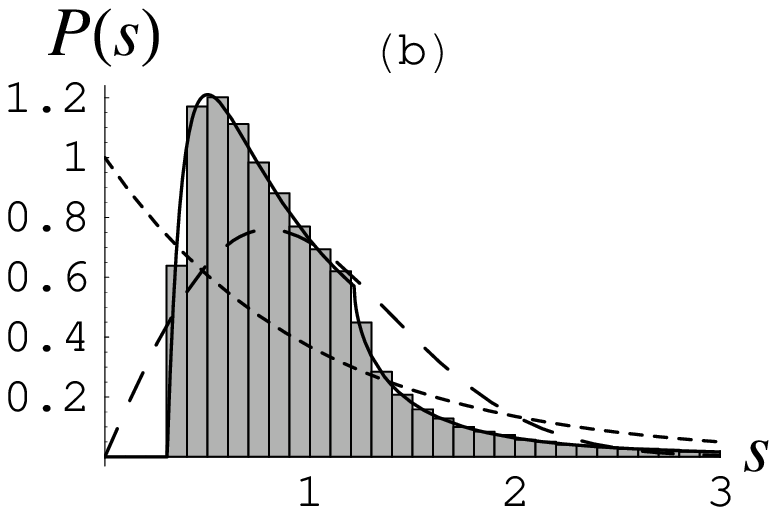} 
    \end{tabular}
    \caption{Separation statistics: (a) for the model Hamiltonian; (b)
    for the Farey sequence.  The solid curves are from the Farey
    spacing measure, \eq{eq:Augustin}.  In (b) the short-dashed curve
    is for the Poisson Process of the generic integrable problem and
    the long-dashed curve is that for the Gaussian orthogonal ensemble
    of random matrices (quantum chaotic case).}
    \label{fig:PP}
\end{figure}

This high degeneracy is also the cause of the delta function spike at
the origin visible in the level-separation probability distribution
plot shown in \fig{fig:PP}(a), which is very similar to Fig.~9(a) in
\cite{dewar_etal04}.

\subsection{Farey statistics}

The tail in \fig{fig:PP}(a) is due to the non-degenerate component of
the spectrum, obtained by reducing all fractions $p/q$ to lowest terms
and deleting duplications.  Thus the eigenvalues in this set are $N_Q$
times the terms of the Farey sequence ${\cal F}(Q)$.

To study the statistics of this non-degenerate component it is natural
to define the \emph{Farey spectrum} $\{E^{\rm F}_i\}$ as $N^{\rm F}(Q)$
times the terms of the Farey sequence ${\cal F}(Q)$, where $N^{\rm
F}(Q)$ is the number of terms in ${\cal F}(Q)$.  The asymptotic
behaviour of $N^{\rm F}(Q)$ in the large-$n$ limit is given \cite[p.
391]{cvitanovich_etal03} by $N^{\rm F}(Q) \sim 3 Q^2/\pi^2 + O(Q\ln Q)$.
The staircase plot and separation distribution $P^{\rm F}(s)$ for the
Farey spectrum are given in \fig{fig:Devil}(b) and \fig{fig:PP}(b),
respectively.

It is a standard result in the theory of Farey sequences 
\cite[p. 301]{niven-zuckerman-montgomery91} that the smallest and 
largest nearest-neighbour spacings in ${\cal F}(Q)$ are given 
respectively by
\begin{equation}
    \frac{1}{Q(Q-1)} \quad \mathrm{and} \quad \frac{1}{Q} \;,
    \label{eq:minmaxspacing}
\end{equation}
so that the support of the tail component of $P(s)$ in \fig{fig:PP}(a)
becomes $[1/2,\infty)$ in the limit $Q \rightarrow \infty$, while 
that of $P^{\rm F}(s)$ in \fig{fig:PP}(b) is $[3/\pi^2,\infty)$.

Augustin \emph{et al.} \cite{augustin_etal01}, eq.  (1.9), derive the
spacing density for the Farey sequence as
\begin{equation}
    g_1(t)  \equiv \left\{ 
    \begin{array}{ll}
	0\,, &
	\mathrm{for} \quad 0 \leq t \leq \frac{3}{\pi^2},  \\
	\frac{6}{\pi^2t^2}\ln\left(\frac{\pi^2t}{3}\right)\,, & 
	\mathrm{for} \quad \frac{3}{\pi^2} \leq t \leq \frac{12}{\pi^2}\,, \\
	\frac{12}{\pi^2t^2}
	\ln \left[\frac{\pi^2t}{6}
		  \left(1 - \sqrt{1 - \frac{12}{\pi^2t}}
		  \right)
	    \right] & 
	\mathrm{for} \quad \frac{12}{\pi^2} \leq t.
    \end{array}
      \right.
    \label{eq:Augustin}
\end{equation}
The solid curve in \fig{fig:PP}(b) is obtained by setting $P^{\rm F}(s) = 
g_1(s)$ and is seen to agree well with the numerical results.

The solid curve in \fig{fig:PP}(a) is obtained by setting $P(s) =
[N^{\rm F}(Q)/N(Q)]^2 g_1(N^{\rm F} s/N)$ and agrees well
with the tail of the histogram.  The ratio of the area of the tail in
\fig{fig:PP}(a) to the strength of the delta function \fig{fig:PP}(a)
is $N^{\rm F}(Q)/[N(Q)-N^{\rm F}(Q)] \approx 1.55$.

We have verified that the probability distributions remain
unchanged if subranges of the spectra are used, in agreement with the
result included in Theorem 1.1 of \cite{augustin_etal01} that the
convergence to a probability measure is independent of the interval
chosen.

\section{W7--X results}
\label{sec:W7X}

\begin{figure}[tbp]
\begin{center}
\begin{tabular}{cc}
      \includegraphics[scale=0.7]{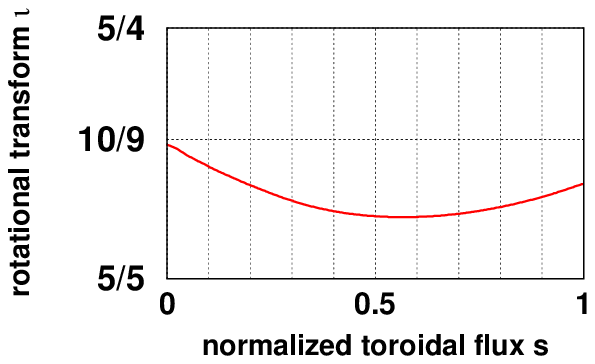}
    & \includegraphics[scale=0.7]{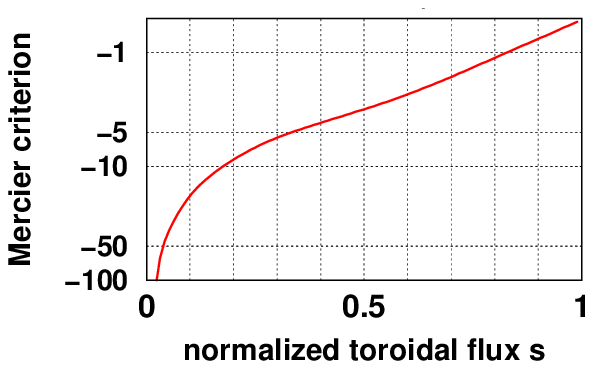}  \\
\end{tabular}
\end{center}
\caption{Left: the rotational transform $\iotabar = \iotabar(s)$
profile versus normalized toroidal flux $s$ ($s \propto r^2$ near the
magnetic axis).  Right: a measure of the Mercier stability versus
normalized toroidal flux for the 3-dimensional W7--X-like case of
\fig{fig:W7X}.  A negative value indicates instability.}
    \label{fig:W7Xprofiles}
\end{figure}

The W7--X variant equilibrium studied was generated with the VMEC
\cite{hirshman-betancourt91} code, which assumes the magnetic field to
be integrable, so that all magnetic field lines lie on nested toroidal
flux surfaces, which we label by $s$, the enclosed toroidal magnetic
flux divided by the toroidal flux enclosed by the plasma boundary.
The magnetic field is characterized on each flux surface by its
winding number $\iotabar(s)$.  (In tokamaks its inverse, $q \equiv
1/\iotabar$, is more commonly used.)  As seen in \fig{fig:W7Xprofiles}
the rotational transform profile is nonmonotonic and has low shear
($\iotabar_\mathrm{axis} = 1.1066$, $\iotabar_\mathrm{min} = 1.0491$,
$\iotabar_\mathrm{edge} = 1.0754$) so it is close to, but greater
than, unity over the whole plasma.  It is also seen from
\fig{fig:W7Xprofiles} equilibrium is interchange unstable, because the
Mercier stability criterion is violated over the whole plasma.

\begin{figure}[tbp]
\begin{center}
    \includegraphics[scale=0.5]{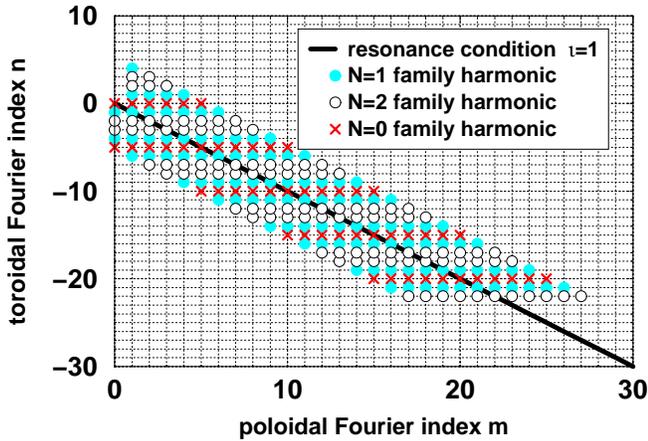}
\end{center}
    \caption{Choice of basis set of toroidal and poloidal Fourier 
    harmonics.}
    \label{fig:mntableau}
\end{figure}

The CAS3D code expands the eigenfunctions in a finite Fourier basis in
the toroidal and poloidal angles, selected so as to include all $|n|
\leq n_{\rm max}$ and all $m$ such that $n/m$ lies in a band including
the range of $\iotabar$.  The Fourier tableau is depicted graphically
in \fig{fig:mntableau}.
In this code,
the radial dependence of the perturbation functions is treated by
a hybrid Finite-Element approach, using a linear interpolation for
the normal displacement and piecewise constant interpolations for
the scalar components that describe the MHD displacement within the
magnetic surfaces.
In the calculations discussed here, 301 radial grid points have been used.
The kinetic energy was used as normalization, and therefore, the unstable
eigenvalues $\lambda$ may be converted to a nondimensional growth rate
$\gamma$ viz.
$\gamma\tau_{\mathrm{A}} = R_0(0) \sqrt{|\lambda |} / B_0(0)$.
Here, $R_0(0)$ is the major radius and $B_0(0)$ the equilibrium
magnetic field measured on the magnetic axis.

Because of the 5-fold symmetry of the equilibrium, any toroidal
Fourier harmonic $n$ in an eigenfunction is coupled to toroidal
harmonics $n \pm 5$. 
With the poloidal harmonics chosen to be positive, $m \geq 0$, there
are just three uncoupled mode families $N = 0,1,2$ 
(compare \cite{schwab93}).

\begin{figure}[tbp]
\begin{center}
    \includegraphics[scale=0.5]{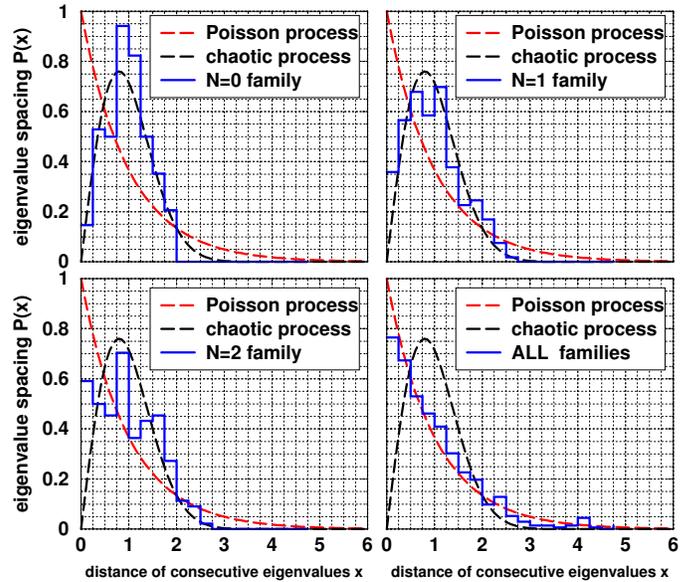}
\end{center}
    \caption{Unfolded eigenvalue spacing distributions from mode
    family datasets $N = 0$ (137 values), $N = 1$ (214 values) and $N
    = 2$ (178 values) calculated by CAS3D for our W7--X-like
    equilibrium, and the distribution for the combined spectrum, $N
    =0, 1$ and $2$.}
    \label{fig:CASPPlot}
\end{figure}

\begin{figure}[tbp]
\begin{center}
    \includegraphics[scale=1]{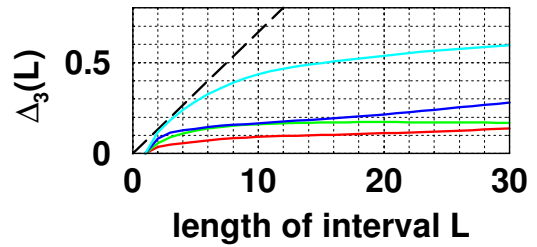}
\end{center}
    \caption{Dyson--Mehta spectral rigidity as a function of 
    subinterval length $L$.
    Colour code: N=0 mode family: red; N=1: green; N=2: blue;
    combined data set: cyan. The Poisson-process limit is also
    indicated (black dashed).}
    \label{fig:Rigidity}
\end{figure}

We characterize the statistics of the ensembles of eigenvalues within
the three mode families using two standard measures from quantum chaos
theory \cite{haake01,mehta91,bruus-dauriac97}, first renormalizing
(``unfolding'') the eigenvalues so their average separation is unity.
The first measure, shown in \fig{fig:CASPPlot}, is the probability
distribution function $P(x)$ for the eigenvalue separation $x$.  The
other, shown in \fig{fig:Rigidity}, is the Dyson--Mehta rigidity
$\Delta_3(L)$, where $L$ is the subrange of unfolded eigenvalues used.

As seen from \fig{fig:CASPPlot}, when
the statistics are analyzed within the three mode families the
eigenvalue spacing distribution function is closer to the Wigner
conjecture form found for generic chaotic systems
\cite{haake01} than to the Poisson distribution for separable systems,
as might be expected from \cite{dewar-cuthbert-ball01}.  However, when the
spectra from the three uncoupled mode familes are combined, there are
enough accidental degeneracies that the spacing distribution becomes
close to Poissonian.

\begin{figure}[tbp]
\begin{center}
    \includegraphics[scale=0.5]{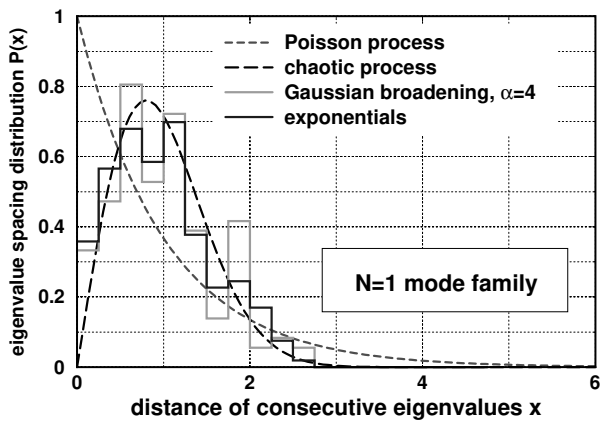}
\end{center}
    \caption{Unfolded eigenvalue spacing distributions for mode family 
    $N = 1$, calculated using  two different unfolding methods. The 
    results are seen to be consistent to within statistical error.}
    \label{fig:CASPPlotN1}
\end{figure}

To test the sensitivity to the precise method of unfolding chosen, we did 
the statistics using two different methods.
They are the Gaussian unfolding \cite{bruus-dauriac97}\ and a fit
with exponentials.
The results, shown 
in \fig{fig:CASPPlotN1}, indicate little sensitivity to unfolding 
method.

\section{Conclusion \label{sec:Conc}}

Although not presented here, when \emph{all} unstable eigenmodes
(i.e. all $l$, $m$, and $n$) are
included, the eigenvalue spacing statistics for the ideal-MHD
interchange eigenvalue spectrum in a separable cylindrical
approximation is close to that of generic separable wave equations
\cite{berry-tabor77}, despite our earlier finding
\cite{dewar-nuehrenberg-tatsuno04,dewar_etal04} that the spectrum in
the subspace of the most unstable radial eigenmode $l = 0$ is 
nongeneric, as explained by the model presented in Sec.~\ref{sec:toy}.

In this paper we have shown that
a strongly three-dimensional stellarator equilibrium related to
W7-X, the unstable interchange (Mercier) mode spectrum has, to within
statistical uncertainties, similar statistics to generic quantum
chaotic systems.  That is, the overwhelming majority of eigenvalues
are not ``good quantum numbers'' and can thus be expected to display
sensitivity to small perturbations.  This needs to be borne in mind
when doing convergence studies using stability codes such as CAS3D.

An interesting question for further work is whether other modes, such
as drift waves, are quantum chaotic in stellarators, or if this is a
pecularity of MHD modes.  There is already evidence that kinetic
effects make the semiclassical (WKB) dynamics closer to integrable 
\cite{mcmillan-dewar06}.

Another question is whether quantum chaos of Mercier modes occurs in 
machines with more field periods than the 5 of W7-X. Earlier work 
\cite{cooper-singleton-dewar96} suggested that, in a 10-field-period 
heliotron equilibrium related to the Large Helical Device (LHD), the 
spectrum is close to that of an equivalent axisymmetric torus, and 
thus not chaotic. However, a relatively few modes were studied and the 
spacing statistics were not calculated. 

\begin{acknowledgments}
One of us (RLD) acknowledges support by the Australian Research Council
and thanks the organizers of the Fourth Dynamics Days Asia Pacific (DDAP4)
conference at which this work was presented. 
\end{acknowledgments}

\bibliography{Ballooning}

\begin{thebibliography}{20}
\expandafter\ifx\csname natexlab\endcsname\relax\def\natexlab#1{#1}\fi
\expandafter\ifx\csname bibnamefont\endcsname\relax
  \def\bibnamefont#1{#1}\fi
\expandafter\ifx\csname bibfnamefont\endcsname\relax
  \def\bibfnamefont#1{#1}\fi
\expandafter\ifx\csname citenamefont\endcsname\relax
  \def\citenamefont#1{#1}\fi
\expandafter\ifx\csname url\endcsname\relax
  \def\url#1{\texttt{#1}}\fi
\expandafter\ifx\csname urlprefix\endcsname\relax\def\urlprefix{URL }\fi
\providecommand{\bibinfo}[2]{#2}
\providecommand{\eprint}[2][]{\url{#2}}

\bibitem[{\citenamefont{Anderson et~al.}(1990)\citenamefont{Anderson, Cooper,
  Gruber, Merazzi, and Schwenn}}]{anderson_etal90}
\bibinfo{author}{\bibfnamefont{D.~V.} \bibnamefont{Anderson}},
  \bibinfo{author}{\bibfnamefont{W.~A.} \bibnamefont{Cooper}},
  \bibinfo{author}{\bibfnamefont{R.}~\bibnamefont{Gruber}},
  \bibinfo{author}{\bibfnamefont{S.}~\bibnamefont{Merazzi}}, \bibnamefont{and}
  \bibinfo{author}{\bibfnamefont{U.}~\bibnamefont{Schwenn}},
  \bibinfo{journal}{Int. J. Supercomp. Appl.} \textbf{\bibinfo{volume}{4}},
  \bibinfo{pages}{34} (\bibinfo{year}{1990}).

\bibitem[{\citenamefont{Schwab}(1993)}]{schwab93}
\bibinfo{author}{\bibfnamefont{C.}~\bibnamefont{Schwab}},
  \bibinfo{journal}{Phys. Fluids B} \textbf{\bibinfo{volume}{5}},
  \bibinfo{pages}{3195} (\bibinfo{year}{1993}).

\bibitem[{\citenamefont{N{\"u}hrenberg}(1996)}]{nuehrenberg96}
\bibinfo{author}{\bibfnamefont{C.}~\bibnamefont{N{\"u}hrenberg}},
  \bibinfo{journal}{Phys. Plasmas} \textbf{\bibinfo{volume}{3}},
  \bibinfo{pages}{2401} (\bibinfo{year}{1996}).

\bibitem[{\citenamefont{Bernstein et~al.}(1958)\citenamefont{Bernstein,
  Frieman, Kruskal, and Kulsrud}}]{bernstein_etal58}
\bibinfo{author}{\bibfnamefont{I.~B.} \bibnamefont{Bernstein}},
  \bibinfo{author}{\bibfnamefont{E.~A.} \bibnamefont{Frieman}},
  \bibinfo{author}{\bibfnamefont{M.~D.} \bibnamefont{Kruskal}},
  \bibnamefont{and} \bibinfo{author}{\bibfnamefont{R.~M.}
  \bibnamefont{Kulsrud}}, \bibinfo{journal}{Proc. R. Soc. London Ser. A}
  \textbf{\bibinfo{volume}{244}}, \bibinfo{pages}{17} (\bibinfo{year}{1958}).

\bibitem[{\citenamefont{Haake}(2001)}]{haake01}
\bibinfo{author}{\bibfnamefont{F.}~\bibnamefont{Haake}},
  \emph{\bibinfo{title}{Quantum Signatures of Chaos}}
  (\bibinfo{publisher}{Springer-Verlag}, \bibinfo{address}{Berlin},
  \bibinfo{year}{2001}), \bibinfo{edition}{2nd} ed.

\bibitem[{\citenamefont{Dewar et~al.}(2004{\natexlab{a}})\citenamefont{Dewar,
  N{\"u}hrenberg, and Tatsuno}}]{dewar-nuehrenberg-tatsuno04}
\bibinfo{author}{\bibfnamefont{R.~L.} \bibnamefont{Dewar}},
  \bibinfo{author}{\bibfnamefont{C.}~\bibnamefont{N{\"u}hrenberg}},
  \bibnamefont{and} \bibinfo{author}{\bibfnamefont{T.}~\bibnamefont{Tatsuno}},
  \bibinfo{journal}{J. Plasma Fusion Res. SERIES} \textbf{\bibinfo{volume}{6}},
  \bibinfo{pages}{40} (\bibinfo{year}{2004}{\natexlab{a}}),
  \bibinfo{note}{proceedings of the 13th International Toki Conference, Toki,
  Japan, 9-12 December 2003}, \eprint{arXiv:physics/0409070}.

\bibitem[{\citenamefont{Dewar et~al.}(2004{\natexlab{b}})\citenamefont{Dewar,
  Tatsuno, Yoshida, N\"uhrenberg, and McMillan}}]{dewar_etal04}
\bibinfo{author}{\bibfnamefont{R.~L.} \bibnamefont{Dewar}},
  \bibinfo{author}{\bibfnamefont{T.}~\bibnamefont{Tatsuno}},
  \bibinfo{author}{\bibfnamefont{Z.}~\bibnamefont{Yoshida}},
  \bibinfo{author}{\bibfnamefont{C.}~\bibnamefont{N\"uhrenberg}},
  \bibnamefont{and} \bibinfo{author}{\bibfnamefont{B.~F.}
  \bibnamefont{McMillan}}, \bibinfo{journal}{Phys. Rev. E}
  \textbf{\bibinfo{volume}{70}}, \bibinfo{pages}{066409}
  (\bibinfo{year}{2004}{\natexlab{b}}), \eprint{arXiv:physics/0405095}.

\bibitem[{\citenamefont{Berry and Tabor}(1977)}]{berry-tabor77}
\bibinfo{author}{\bibfnamefont{M.~V.} \bibnamefont{Berry}} \bibnamefont{and}
  \bibinfo{author}{\bibfnamefont{M.}~\bibnamefont{Tabor}},
  \bibinfo{journal}{Proc. R. Soc. Lond. A} \textbf{\bibinfo{volume}{356}},
  \bibinfo{pages}{375} (\bibinfo{year}{1977}).

\bibitem[{\citenamefont{Hameiri}(1985)}]{hameiri85}
\bibinfo{author}{\bibfnamefont{E.}~\bibnamefont{Hameiri}},
  \bibinfo{journal}{Commun. Pure Appl. Math.} \textbf{\bibinfo{volume}{38}},
  \bibinfo{pages}{43} (\bibinfo{year}{1985}).

\bibitem[{\citenamefont{Niven et~al.}(1991)\citenamefont{Niven, Zuckerman, and
  Montgomery}}]{niven-zuckerman-montgomery91}
\bibinfo{author}{\bibfnamefont{I.}~\bibnamefont{Niven}},
  \bibinfo{author}{\bibfnamefont{H.~S.} \bibnamefont{Zuckerman}},
  \bibnamefont{and} \bibinfo{author}{\bibfnamefont{H.~L.}
  \bibnamefont{Montgomery}}, \emph{\bibinfo{title}{An Introduction to the
  Theory of Numbers}} (\bibinfo{publisher}{Wiley}, \bibinfo{address}{New York},
  \bibinfo{year}{1991}), \bibinfo{edition}{5th} ed.

\bibitem[{\citenamefont{Artuso et~al.}(1989)\citenamefont{Artuso, Cvitanovi\'c,
  and Kenny}}]{artuso-cvitanovic-kenny89}
\bibinfo{author}{\bibfnamefont{R.}~\bibnamefont{Artuso}},
  \bibinfo{author}{\bibfnamefont{P.}~\bibnamefont{Cvitanovi\'c}},
  \bibnamefont{and} \bibinfo{author}{\bibfnamefont{B.~G.} \bibnamefont{Kenny}},
  \bibinfo{journal}{Phys. Rev. A} \textbf{\bibinfo{volume}{39}},
  \bibinfo{pages}{268} (\bibinfo{year}{1989}).

\bibitem[{\citenamefont{Augustin et~al.}(2001)\citenamefont{Augustin, Boca,
  Cobeli, and Zaharescu}}]{augustin_etal01}
\bibinfo{author}{\bibfnamefont{V.}~\bibnamefont{Augustin}},
  \bibinfo{author}{\bibfnamefont{F.~P.} \bibnamefont{Boca}},
  \bibinfo{author}{\bibfnamefont{C.}~\bibnamefont{Cobeli}}, \bibnamefont{and}
  \bibinfo{author}{\bibfnamefont{A.}~\bibnamefont{Zaharescu}},
  \bibinfo{journal}{Math. Proc. Camb. Phil. Soc.}
  \textbf{\bibinfo{volume}{131}}, \bibinfo{pages}{23} (\bibinfo{year}{2001}).

\bibitem[{\citenamefont{Abbott}(1992)}]{abbott92}
\bibinfo{author}{\bibfnamefont{P.}~\bibnamefont{Abbott}}, \bibinfo{journal}{The
  Mathematica Journal} \textbf{\bibinfo{volume}{2}} (\bibinfo{year}{1992}),
  \bibinfo{note}{http://www.mathematica-journal.com/issue/v2i2/}.

\bibitem[{\citenamefont{Cvitanovi\'c et~al.}(2003)\citenamefont{Cvitanovi\'c,
  Artuso, Mainieri, Tanner, and Vattay}}]{cvitanovich_etal03}
\bibinfo{author}{\bibfnamefont{P.}~\bibnamefont{Cvitanovi\'c}},
  \bibinfo{author}{\bibfnamefont{R.}~\bibnamefont{Artuso}},
  \bibinfo{author}{\bibfnamefont{R.}~\bibnamefont{Mainieri}},
  \bibinfo{author}{\bibfnamefont{G.}~\bibnamefont{Tanner}}, \bibnamefont{and}
  \bibinfo{author}{\bibfnamefont{G.}~\bibnamefont{Vattay}},
  \emph{\bibinfo{title}{Classical and Quantum Chaos}} (\bibinfo{publisher}{{\tt
  ChaosBook.org} Niels Bohr Institute}, \bibinfo{address}{Copenhagen},
  \bibinfo{year}{2003}), \bibinfo{edition}{10th} ed., \bibinfo{note}{webbook:
  http://chaosbook.org/}.

\bibitem[{\citenamefont{Hirshman and Betancourt}(1991)}]{hirshman-betancourt91}
\bibinfo{author}{\bibfnamefont{S.~P.} \bibnamefont{Hirshman}} \bibnamefont{and}
  \bibinfo{author}{\bibfnamefont{O.}~\bibnamefont{Betancourt}},
  \bibinfo{journal}{J. Comput. Phys.} \textbf{\bibinfo{volume}{96}},
  \bibinfo{pages}{99} (\bibinfo{year}{1991}).

\bibitem[{\citenamefont{Mehta}(1991)}]{mehta91}
\bibinfo{author}{\bibfnamefont{M.~L.} \bibnamefont{Mehta}},
  \emph{\bibinfo{title}{Random Matrices}} (\bibinfo{publisher}{Academic Press},
  \bibinfo{address}{San Diego}, \bibinfo{year}{1991}), \bibinfo{edition}{2nd}
  ed.

\bibitem[{\citenamefont{Bruus and d'Auriac}(1997)}]{bruus-dauriac97}
\bibinfo{author}{\bibfnamefont{H.}~\bibnamefont{Bruus}} \bibnamefont{and}
  \bibinfo{author}{\bibfnamefont{J.-C.~A.} \bibnamefont{d'Auriac}},
  \bibinfo{journal}{Phys. Rev. B} \textbf{\bibinfo{volume}{55}},
  \bibinfo{pages}{9142} (\bibinfo{year}{1997}).

\bibitem[{\citenamefont{Dewar et~al.}(2001)\citenamefont{Dewar, Cuthbert, and
  Ball}}]{dewar-cuthbert-ball01}
\bibinfo{author}{\bibfnamefont{R.~L.} \bibnamefont{Dewar}},
  \bibinfo{author}{\bibfnamefont{P.}~\bibnamefont{Cuthbert}}, \bibnamefont{and}
  \bibinfo{author}{\bibfnamefont{R.}~\bibnamefont{Ball}},
  \bibinfo{journal}{Phys. Rev. Letters} \textbf{\bibinfo{volume}{86}},
  \bibinfo{pages}{2321} (\bibinfo{year}{2001}), \bibinfo{note}{e-Print
  arXiv:physics/0102065}.

\bibitem[{\citenamefont{McMillan and Dewar}(2006)}]{mcmillan-dewar06}
\bibinfo{author}{\bibfnamefont{B.~F.} \bibnamefont{McMillan}} \bibnamefont{and}
  \bibinfo{author}{\bibfnamefont{R.~L.} \bibnamefont{Dewar}},
  \bibinfo{journal}{Nucl. Fusion} \textbf{\bibinfo{volume}{46}},
  \bibinfo{pages}{477} (\bibinfo{year}{2006}).

\bibitem[{\citenamefont{Cooper et~al.}(1996)\citenamefont{Cooper, Singleton,
  and Dewar}}]{cooper-singleton-dewar96}
\bibinfo{author}{\bibfnamefont{W.~A.} \bibnamefont{Cooper}},
  \bibinfo{author}{\bibfnamefont{D.~B.} \bibnamefont{Singleton}},
  \bibnamefont{and} \bibinfo{author}{\bibfnamefont{R.~L.} \bibnamefont{Dewar}},
  \bibinfo{journal}{Phys. Plasmas} \textbf{\bibinfo{volume}{3}},
  \bibinfo{pages}{275} (\bibinfo{year}{1996}), \bibinfo{note}{erratum: Phys.
  Plasmas \textbf{3}, 3520 (1996)}.

\end{thebibliography}
\end{document}